\documentclass[aps,amsmath,amssymb,twocolumn,showpacs]{revtex4}
\usepackage{ulem}
\usepackage{color}

\usepackage{verbatim}
\usepackage{graphicx}
\usepackage{bm}
\def\pmb#1{\setbox0=\hbox{#1}
\kern-.025em\copy0\kern-\wd0 \kern-.05em\copy0\kern-\wd0
\kern-.025em\raise.0433em\box0}

\newcommand{\beq}{\begin{equation}}
\newcommand{\eeq}{\end{equation}}
\newcommand{\ba}{\begin{eqnarray}}
\newcommand{\ea}{\end{eqnarray}}

\input epsf

\newif\iffigures

\figurestrue 

\begin{document}

\title[]{Acoustic flat lensing using an indefinite medium}

\author{M. Dubois$^{1}$, J. Perchoux$^{2}$, A.~L. Vanel$^{3}$, C. Tronche$^{2}$, Y. Achaoui$^{4}$, G. Dupont$^{5}$, K. Bertling$^{6}$, A.~D. Raki\'c$^{6}$, T. Antonakakis$^{7}$,
S. Enoch$^{1}$, R. Abdeddaim$^{1}$, R.~V. Craster$^{3}$, S. Guenneau$^{1}$}
\affiliation{$^1$ Aix Marseille Univ, CNRS, Centrale Marseille, Institut Fresnel, Marseille, France}
\affiliation{$^2$ LAAS-CNRS, Universit\'e de Toulouse, CNRS, INP, Toulouse, France}
\affiliation{$^3$ Department of Mathematics, Imperial College London, London SW7 2AZ, UK}
\affiliation{$^4$ Universit\'e de Franche-Comt\'e, CNRS, ENSMM, FEMTO-ST, 25000 Besancon, France} 
\affiliation{$^5$ Aix$-$Marseille Univ, CNRS, Centrale Marseille, IRPHE, Marseille, France}
\affiliation{$^6$   School of Information Technology
and Electrical Engineering, The University of Queensland, Brisbane, 4072, Australia}
\affiliation{$^7$ Multiwave Technologies AG, 3 Chemin du Pr\'e Fleuri, 1228 Geneva, Switzerland}
\begin{abstract}

Acoustic flat lensing is achieved here by tuning a phononic array to have indefinite medium behaviour in a narrow frequency spectral region along the acoustic branch. This is  confirmed by the occurrence of a flat band along an unusual path in the Brillouin zone and by interpreting the intersection point of isofrequency contours on the corresponding isofrequency surface; coherent directive beams are formed whose reflection from the array surfaces create lensing. Theoretical predictions are corroborated by time-domain experiments, airborne acoustic waves generated by a source with a frequency centered about $10.6$ kHz, placed at three different distances 
 from one side of a finite phononic crystal slab, constructed from
 polymeric spheres, yield distinctive focal spots on the other side.  
 These experiments evaluate the pressure field using optical feedback interferometry and demonstrate precise control of the three-dimensional wave trajectory through a sonic crystal. 


\end{abstract}
\maketitle

\begin{figure}
\centering
\includegraphics[width=7cm]{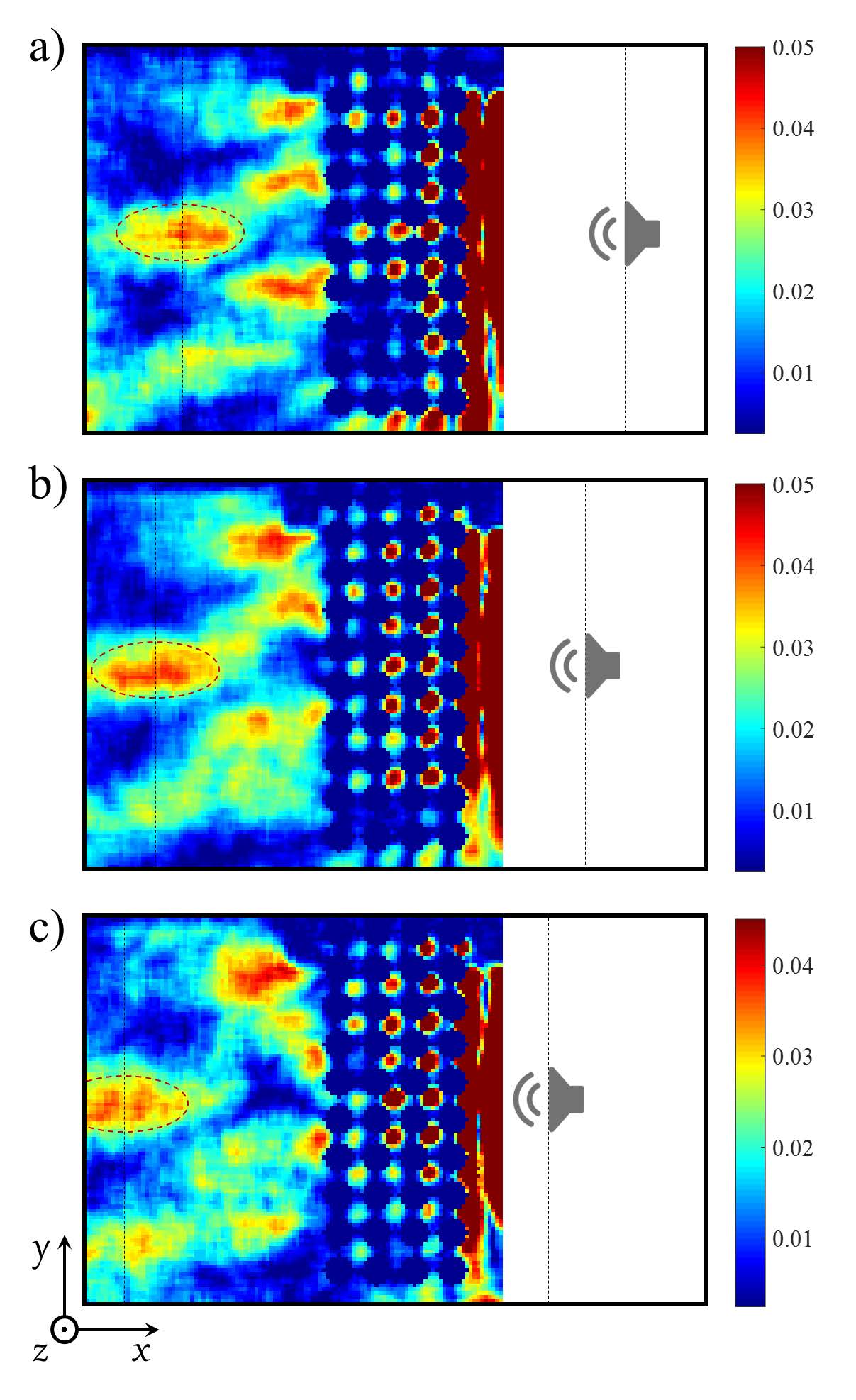}
\caption{Envelope of pressure amplitude measured at time $t=1.1$ ms for three source positions: (a) $6.4$ cm; (b) $4.8$ cm; (c) $3.2$ cm; Wavelength in air is $3.2$cm at $10.3$ kHz for a sound velocity of 330~m/s. Schematics show the source position (not in the field of view) and a dotted ellipsoid surrounds the focal spot on the left side of the lens. Linear color scale is in arbitrary units (same calibration for the three experiments). See Supplemental Material at [URL] for a movie showing the dynamics of the lens for $t\in[0.624,2.176]$ ms.}
\label{fig0}
\end{figure}

The band spectra of photonic \cite{sakoda01a,joannopoulos08a} and phononic \cite{kushwaha93a} crystals can be interpreted to predict a rich array of interesting physical effects, for instance anomalous refraction \cite{sakoda01a,gralak00a} and all-angle-negative refraction \cite{luo02a} amongst many others; understanding these spectra underpins advances in electronic properties, wave transport in photonics and acoustics, as well as in interference phenomena throughout many fields of physical and engineering sciences. 

In this Letter we report experimental results where an image of a volumetric source through a three-dimensional phononic crystal (PC) forms according to the physics of indefinite media \cite{smith03a}, see Fig.~\ref{fig0}. The image is not created by tilting the crystal as in acoustic superlenses \cite{page04}, or at low frequencies using effective media \cite{cervera02} nor by negative refraction acoustic flat lenses using metamaterials \cite{zhang09,kaina2015}. Instead we identify critical points on the isofrequency surfaces, for a simple cubic array of rigid spheres, where beam-like trajectories are formed and use these beams, and their reflections, to create lensing; this is using the properties of indefinite media \cite{smith03a}.

The first experimental demonstration of a three-dimensional flat acoustic lens in 2004 \cite{page04} used 0.8-mm tungsten carbide beads surrounded by water, with the beads closely packed in a face-centered cubic crystal structure along the body diagonal ($\Gamma R$ crystal direction); the lensing function  was above the phononic band gap of the PC and at $1.57$ MHz with the pressure waves focused into a tight spot (about $5$~mm). We give an alternative design to this well-known phononic lens, based upon a different physical mechanism, also exhibiting focusing reminiscent of the Veselago-Pendry convergent flat lens \cite{veselago68a,pendry00a}, see Fig.~\ref{fig0}.
Negative refraction and superlensing of underwater pressure waves has been theorized with an anisotropic acoustic metamaterial formed by layers of perforated rigid plates \cite{christensen12}, that is analogous to a hyperbolic medium in electromagnetism \cite{belov13a}.

 As in \cite{page04} we use an array of sound-hard spheres, although now in air, take a primitive cubic array, 
  and take advantage of recent advances in the imaging of pressure waves by optical feedback interferometry (OFI) \cite{bertling_OE_2014} to verify our predictions experimentally. 
This methodology was developed to perform pressure wave imaging through the monitoring of the refractive index changes in transparent media and we show it provides a versatile and adjustable tool to address complex acoustic wavefields in structured media such as PCs; the OFI system has advantages  over optical imaging systems based on opto-acoustic effect \cite{zipser_AO_2003,malkin_JSV_2014} of compactness and simplicity of optical configuration. Since we operate in air the frequencies are orders of magnitude lower than those of \cite{page04}, 10.6 kHz, and we take a cubic array of 40 polymer spheres $1.38$ cm in diameter with a center-to-center spacing $a=1.5$ cm. We operate below the band-gap, on the acoustic branch, but not at low frequencies where conventional long-wavelength effective media approximations hold. 
 
\begin{figure}[ht!]
\includegraphics[width=8.6cm]{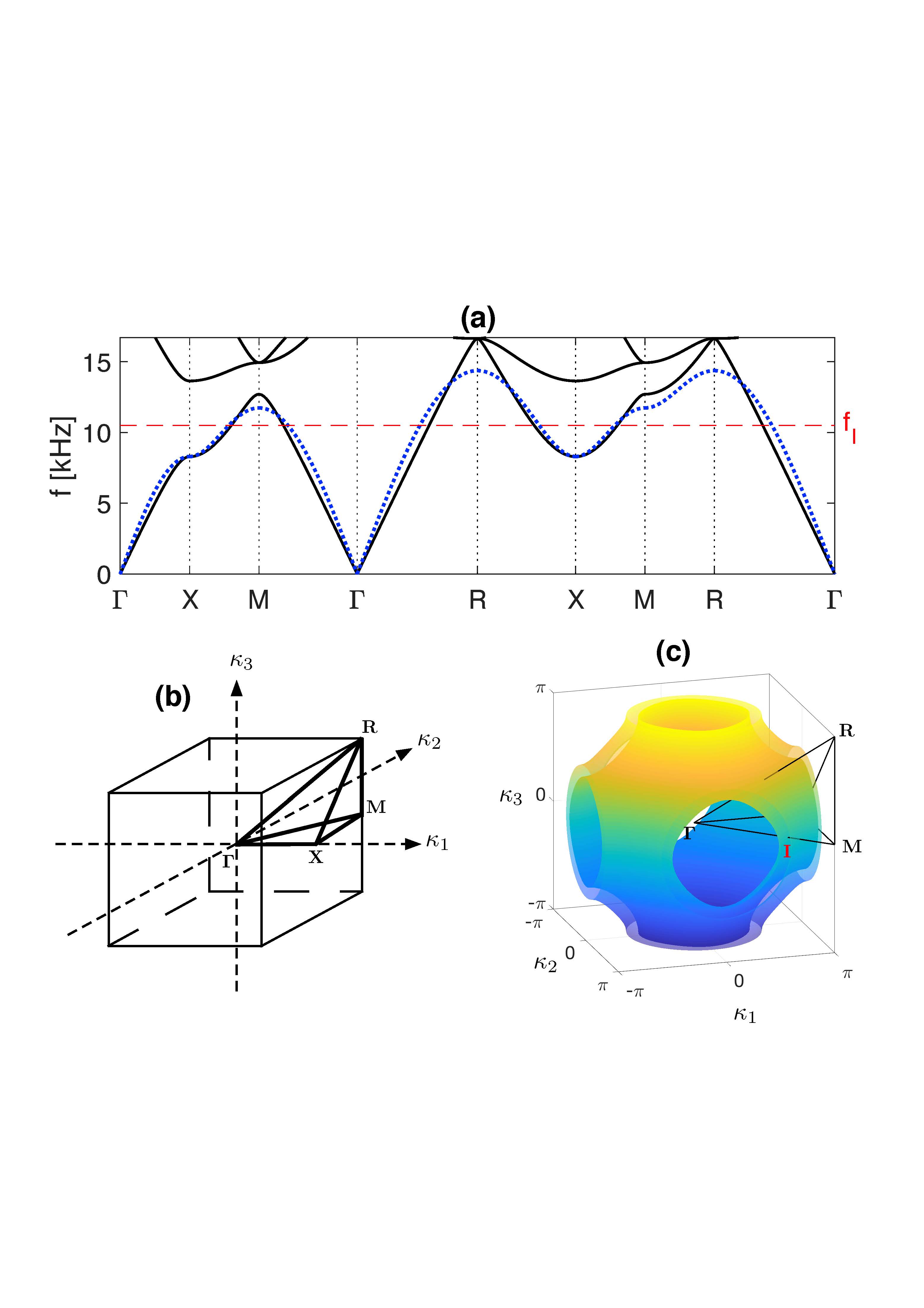}
\caption{Dispersion curves and isofrequency surfaces: (a) Band diagram for the continuum model (black solid) of a cubic array of rigid spheres, $1.38$ cm in diameter with a center-to-center spacing $d=1.5$ cm, versus the discrete spring-mass model (blue dotted) with red dashed line showing frequency $f_I=10.6$ kHz. (b) Superimposed isofrequency surfaces for both continuous (transparent outer surface) and discrete (inner surface) models at frequency $f_I$, with position of the critical point I shown. (c) IBZ used in (a).
}
\label{fig1}
\end{figure}

We aim to gain physical insight and design the crystal by drawing upon simple discrete mass-spring models. Powerful numerical methods, e.g. the plane-wave and multipole expansions, have been
developed that solve the Schr\" odinger \cite{wang1994}, Maxwell \cite{soukoulis90,johnson2001} and Navier \cite{kushwaha93a,poulton2000} equations, and other popular
numerical algorithms include finite elements and finite difference time domain methods: All of these methods require heavy computational resources, particularly in three dimensions which tends to limit research advances in triply periodic media, such as the crystal we consider here. Perhaps remarkably one observes that the isofrequency surfaces (and not merely the dispersion curves along the edges of the Brillouin zone) found numerically for the phononic crystal, for the acoustic branch, are almost identical to those from a mass-spring system, see Fig. \ref{fig1}(a) that shows these surfaces overlain. Moreover, for the mass-spring system highly directive anisotropy occurs at critical points in the Brillouin zone and these are associated with effective anisotropic media \cite{vanel16a}, akin to indefinite media \cite{belov13a}. Given the striking similarity of the isofrequency surfaces vis-\`a-vis the discrete and continuous models we draw conclusions from the discrete model and transfer them to the continuum model thereby avoiding large numbers of expensive simulations. 

We also highlight that identifying the critical points, and the nature of the modes, responsible for the focusing effect, from the standard band diagram, see Fig.~\ref{fig1}(a), going around the edges of irreducible Brillouin Zone (IBZ) is not sufficient for accurate interpretation or identification of frequencies of interest. The complexity of the complete band structure of a 3D phononic crystal (PC) is only appreciated by looking at isofrequency surfaces, see Fig. \ref{fig1}(b): The full dispersion surfaces live in  four-dimensions, so cannot be plotted as such.

For the physical model we use the acoustic pressure field, $p$, that satisfies the wave equation: 
\(
\ddot p=c^2\nabla^2 p
\), where $c$ is the sound speed in air (taken as $330$~m/s), $\nabla^2$ is the spatial Laplacian and the dot decoration denotes time differentiation. We operate in the frequency domain where $f$ is the wave frequency in Hertz. 

\begin{figure}[h!]
\includegraphics[width=8.6cm]{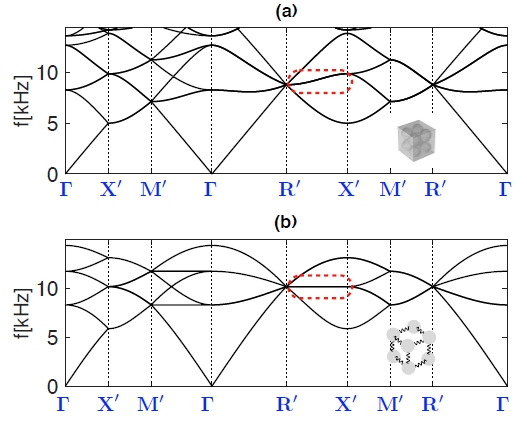}
\vspace{-3.5cm}
\caption{Supercell dispersion curves in the folded IBZ $(\Gamma X'M'R')$ for (a) the continuum model of $8$ spheres 
 and (b) the discrete lattice approximation of $8$ masses: A flat mode highlighted by red dashed rectangle appears between $R' X'$ that is absent in the conventional path Fig.~\ref{fig1}(a). 
The folded Brillouin zone is shown in blue in Fig.~\ref{fig1}(c). 
}
\label{fig2}
\end{figure}

We begin our analysis by considering an infinite phononic crystal, and invoking Bloch's theorem \cite{kittel96a,brillouin53a} to consider a single cubic cell of sidelength $a$, containing a sound-hard sphere (the polymeric spheres are effectively rigid) with Bloch conditions applied to the cell faces. The Bloch wave-vector $\bm\kappa=(\kappa_1,\kappa_2,\kappa_3)$
 characterizes the phase-shift going from one cell to the
 next and dispersion curves for the continuum case in Figs.~\ref{fig1},\ref{fig2} are 
computed with finite element (FE) methods using Comsol Multiphysics, the discrete analogue is a three-dimensional mass-spring lattice of identical masses placed upon a cubic lattice, the dispersion relation is explicit:
\beq
f\sim (f_X/\sqrt{2})
\sqrt{3-\cos(a\kappa_1)-\cos(a\kappa_2)-\cos(a \kappa_3)} .
\label{eq:spring}
\eeq
At point $X$ the standing
wave frequency of the continuum model is $f_X$ which we use to provide the comparison in Fig.~\ref{fig1}(a); it is remarkable that the acoustic branch of the continuum acoustic model is captured so well by this simple discrete model (to within a multiplicative rescaling).

We give both continuum and discrete dispersion curves and construct two sets of  curves: The conventional ones using the standard IBZ (Fig. \ref{fig1}(a) $\Gamma XMR$) and a second set using a supercell which highlights a flatband that is not seen in the conventional approach and for which we use the path $\Gamma X'M' R'$ from Fig.~\ref{fig1}(c). 
 
At first sight it is not clear that there is any advantage in using a supercell and folding the Brillouin zone, the conventional path can under some rather exotic circumstances     miss important details such as the stop band minima/maxima not occuring at the edges of the Brillouin zone
\cite{harrison07a,adams08a}, but this is not our current focus. Instead we note that one can, in 2D,  miss flat bands inside the Brillouin zone that lead to strong
anisotropy \cite{craster12a}. We will operate at 10.6 kHz which looks a completely innocuous frequency in the conventional band diagram of Fig.~2(a), but in the folded band-diagram of Fig.~3(a,b) we see a nearly-flat or completely  (for continuous and discrete cases respectively) flat band connecting $R'X'$ at that frequency. Further exploring this frequency, we show isofrequency surfaces for the discrete and continuous cases in Fig.~2(b) with the flat lines unfolded, where they form square contours, and placed upon the surface. It is particularly notable that these squares intersect at 8 points, one of which we label $I$ for future reference, and for which the direction $\Gamma I$ points along $\Gamma R$ to the corners of the Brillouin zone. This point $I$ is where three lines cross, and for which the group velocity directed along those lines is zero, clearly the group velocity itself is not completely zero but this intersection creates a critical point and energy is preferentially directed along $\Gamma I$; the discrete theory is given by \cite{vanel16a}. Since the full isofrequency surfaces are captured by the discrete  model, see Fig.~2(b), then by extension so is the physics. With this insight we could simply use the discrete model henceforth, but we also  computed full FE simulations for the continuum model at the frequency $f_I=10.6$ kHz that we have identified. Computations of large finite cubic arrays of spheres show that indeed much of the energy is directed to the corners, along the path predicted, and concentrated rays form (see \cite{vanel16a} for discrete computations). These rays are reflected from the faces, of the large finite experimental array, perpendicular to the source and then refocus at the other side as seen in Fig.~\ref{fig0}.

The strongly anisotropic directionality of the highly concentrated rays is suggestive that the underlying character of the equations has changed from elliptic to hyperbolic, with the rays being characteristics. This interpretation is confirmed using high frequency homogenisation \cite{craster10a} to generate an effective medium equation characterised by a tensor that shows, in frequency, when the equations become hyperbolic. The long-scale pressure envelope field, $P$, satisfies 
\beq
{\bf T}\nabla^2 P
-({f}^2-{f_I}^2)P=0,
\eeq
where ${\bf T}$ is a diagonal matrix and we see immediately that entries $T_{11},T_{22},T_{33}$ with the 
 same sign leads to an elliptic equation, conversely opposite signs
 leads to a hyperbolic equation. We draw upon \cite{vanel16a} where the discrete effective medium is created, the coefficients at frequency $f_I$ have  $T_{11}=T_{22}=-8.6$ and $T_{33}=17.2$, showing 
the effective medium to have indefinite medium 
behaviour at frequency $f_I$. We note that in \cite{vanel16a}, such an effective medium is termed hyperbolic medium, but it differs from the physics described in \cite{belov13a}.

\begin{figure}
\centering
\includegraphics[width=8cm]{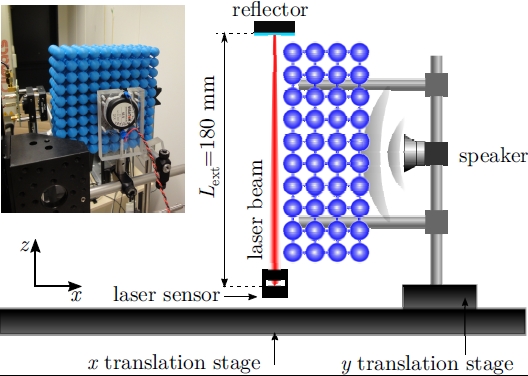}
\caption{Top-view of the experimental setup with photograph as inset. The interferometric sensor consists of a commercial laser diode with a packaged monitoring photodiode, focusing optics, custom made laser driver and signal amplification circuits. Raster scanning uses two long-range translation stages. 
}
\label{fig3}
\end{figure}

To validate these model predictions, the experimental PC, as shown in the inset of Fig.~\ref{fig3}, was built using $1.38$ cm-diameter polylactid polymer spheres 
 machined with a 3D printer and connected to form a 10 $\times$ 10 $\times$ 4 cubic array with 1.5~cm lattice spacing. Each sphere is
attached to its neighbors by 6 small cylinders $0.2$ cm in diameter.

The pressure field is measured point-by-point with a broadband interferometric laser probe based on the OFI sensing scheme \cite{Bosch_EOS_05,taimre_AOP_15}. In this configuration, the laser light is emitted towards a distant target and is partially back-reflected towards the laser cavity where it produces interferences with the inner cavity light. These intra-cavity interferences generate variations of the laser emitted power that can be recorded using any photodetector or directly by monitoring the laser diode voltage \cite{AlRoumy_AO_15}. The pressure variations are sensed using the opto-acoustic effect that induces changes of the refractive index \cite{scruby1990} and thus of the optical path between the laser diode and the reflector in the so-called external cavity where the sound wave propagates. Bertling \textit{et al.} \cite{bertling_OE_2014}, who first proposed this measurement technique, stated that, under the condition that the optical path change remains weak with regards to the laser half-wavelength, the variation of the laser power ${\cal P}(t)$ follows a simple relationship with the refractive index variation such as

\begin{equation}
{\cal P}(t)={\cal P}_0 \cos\left(2\pi f\int_0^L{\frac{2\delta n(z,t)}{c}{\rm d}z} + \phi\right),
\label{eq:ofi}
\end{equation}

\noindent where ${\cal P}_0$ is the power variation amplitude, $L$ is the length of the external cavity (i.e. the distance from the laser to the reflector), $f$ is the laser frequency, $\delta n(z,t)$ is the variation of the refractive index, $c$ is the celerity of light in vacuum and $\phi$ is a constant phase term.\\

\begin{figure}
\centering
\includegraphics[width=8cm]{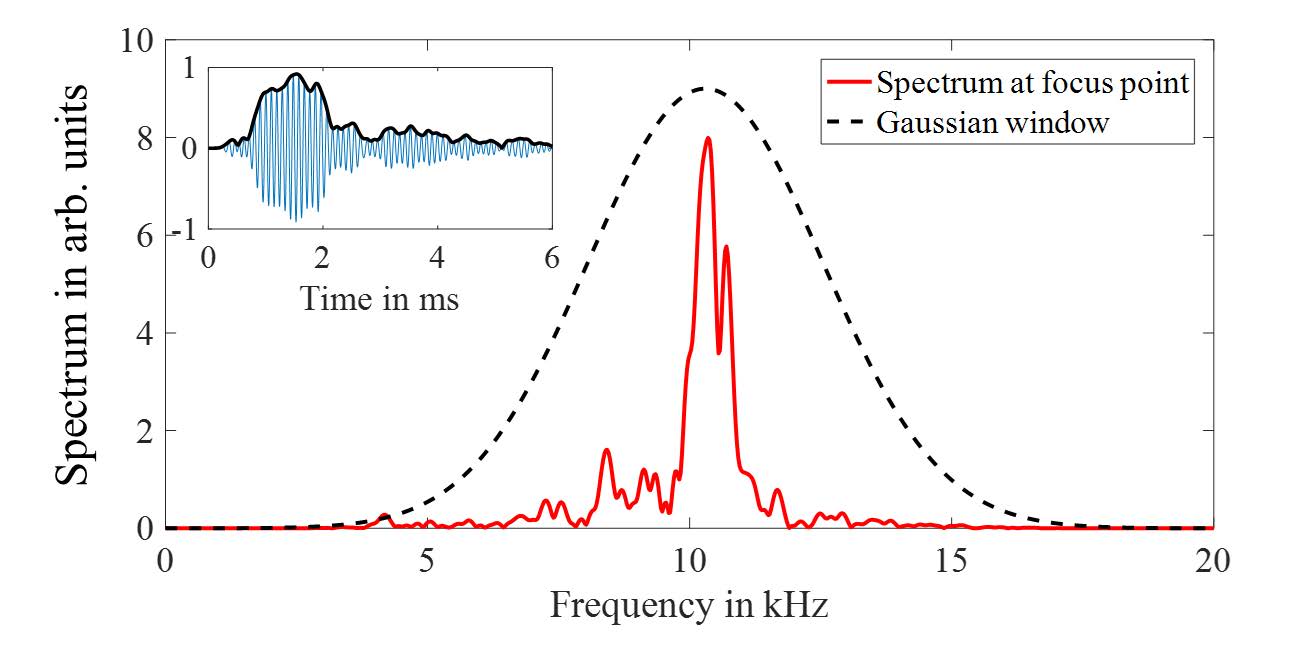}
\caption{Experimental measurements: Spectrum at the focus point (red solid line) for a source located at 6.4 cm from the lens. Black dotted line shows the Gaussian window used to filter the time-dependent signal (centered at $10.3$ kHz, FWHM of $5.2$ kHz). The normalized filtered time trace is shown in the inset. 
}
\label{fig4}
\end{figure}

With the configuration deployed for the present Letter, the reflector consists of a rigid metallic surface covered by a retro-reflective tape and the sensor is a commercial laser diode emitting a single wavelength of 1310~nm associated with an in-package monitoring photodiode whose amplified current is the sensor signal. An aspheric lens focusses the laser beam on the reflector located at 180~mm so that the phononic crystal structure of total length 150~mm fits in-between the sensor and the reflector. The speaker is driven by a function generator producing bursts with 12 periods of a sinusoidal signal at 10.6~kHz and with a maximum achievable power of 96~dBA. Under these conditions, the changes of the optical path in the external cavity are much less than the half wavelength of the laser diode and the sensor signal is an image of the changes of refractive index integrated along the light round-trip in the external cavity  \cite{bertling_OE_2014}.

The loudspeaker and crystal are mounted on a metallic rod assembly, so that the propagation axis of the pressure wave is perpendicular to the crystal surface at its center. The assembly is moved along a 150~mm~$\times$~210~mm grid in steps of 1.5~mm using two long range translation stages while the interferometer remains in a fixed position. To reconstruct the spatio-temporal distribution of the pressure field, each measurement is synchronized with the function generator signal for phase reference. The acoustic burst generation, the scanning displacement and the acquisition with a sampling rate of 1~MS/s are controlled using a National Instrument multifunction data acquisition card. The data is spectrally filtered to obtain the acoustic response to a narrower Gaussian pulsed excitation with chosen central frequency and bandwidth. This post processing allows us to explore the spectral region of interest in the band diagram and validate the discrete spring-mass model predictions. In addition, a median filter is applied to remove spatial coarse noise. The spectrum at the focal spot on the other side of the lens is presented in Fig.~\ref{fig4}, and it shows  maximum transmitted power around 10.3~kHz. The inset shows the time trace at the same location once the Gaussian window centered at 10.3~kHz is applied to the signal. 
This experimental observation of the focusing effect confirms the theoretical prediction of the dynamic anisotropy of the rigid sphere medium leading to the physics of indefinite media \cite{smith03a}.

One interesting aspect of the focusing effect through an indefinite medium is the specific conjugation relation between the source and the focal point. As for negative index lenses, one expects to conserve the distance between the source and the image while moving the source closer to the lens \cite{pendry00a,veselago68a}. We investigate this effect by repeating the previous experiment with a point source located at $4.8$ cm (1.5 wavelength) and $3.2$~cm (1 wavelength) away from the sphere lattice. Fig.~\ref{fig0} presents three snapshots of the envelope of the pressure amplitude at time  $t=1.1$~ms for the three source positions. Identical Gaussian filtering centered at 10.3 kHz is applied to every measurement. A dotted ellipsoid is superimposed on the results to denote the position of the focal spot. A schematic depicts the position of the source in every case. One can clearly observe that the position of the focal spot is moved further away from the sphere lattice as the source is brought closer. Moreover, the distance between source and image is almost kept constant in the three experiments. This distance corresponds to three times the thickness of the lens ($\pm4\%$). Finally, it is important to point out that in each of these experiments, the lateral resolution of the focal spot verifies the Abbe diffraction limit. This observation confirms the connection between the focusing effect observed through this rigid sphere lattice and the hyperbolic behavior predicted by the dispersion relation.

We also conducted time-harmonic FE computations for the Helmholtz equation for sources of central frequencies $9.5$, $10$ and $10.5$ kHz which are in excellent agreement with experimental results for sources at frequencies $9$, $10$ and $11$ kHz (with a bandwidth of $1$ kHz), and confirm `X'-shape emissions at these frequencies, respectively. The offset with the theoretically predicted frequencies and experiments is due to the simple acoustic model that assumes rigid spheres approximating full elasticity equation within the spheres and is described in Fig.~\ref{fig4} and also due to the finite nature of the experiment and unwanted unavoidable minor reflections from the edges of the experimental PC. 

In conclusion, following an earlier theoretical investigation of dynamic anisotropy for a discrete mass-spring cubic lattice \cite{vanel16a}, acoustic pressure waves, interacting with an array of solid spheres surrounded by air, are shown to have highly anisotropic directivity for a source located within the array, and a lensing effect for a source placed outside, thereby opening new avenues in hyperbolic-type metamaterial physics \cite{belov13a} for acoustic waves. Experimental results are in excellent agreement with numerical and asymptotic results. Moreover, such designs could be scaled up or down in order to achieve some control of lower or higher frequency pressure waves. 
We observe a focusing effect reminiscent of what one might get in 2D with all-angle negative refraction \cite{luo02a} by tilting an array, or in 3D \cite{page04} where again the array is tilted, i.e. orientated as a face-centered cubic crystal. Here we take advantage of a critical point, that is hidden in the usual IBZ dispersion curves, that provides highly directional energy propagation along rays that means there is no need to tilt the array to observe lensing. This highly directional behaviour is related to a radical change of character of the underlying effective equation from elliptic to hyperbolic and this exemplifies both the high degree of wave control available in phononic crystals and the importance of a simple and effective predictive model.

R.V.C., A.L.V. and S.G. thank the EPSRC (UK) for support through research grant number EP/J009636/1. K.B. and A.D.R. thank the Australian Research Council (ARC) Discovery Projects funding scheme (DP 160 103910).

\bibliographystyle{apsrev}
\bibliography{references}

\begin{thebibliography}{32}
\expandafter\ifx\csname natexlab\endcsname\relax\def\natexlab#1{#1}\fi
\expandafter\ifx\csname bibnamefont\endcsname\relax
  \def\bibnamefont#1{#1}\fi
\expandafter\ifx\csname bibfnamefont\endcsname\relax
  \def\bibfnamefont#1{#1}\fi
\expandafter\ifx\csname citenamefont\endcsname\relax
  \def\citenamefont#1{#1}\fi
\expandafter\ifx\csname url\endcsname\relax
  \def\url#1{\texttt{#1}}\fi
\expandafter\ifx\csname urlprefix\endcsname\relax\def\urlprefix{URL }\fi
\providecommand{\bibinfo}[2]{#2}
\providecommand{\eprint}[2][]{\url{#2}}

\bibitem[{\citenamefont{Sakoda}(2001)}]{sakoda01a}
\bibinfo{author}{\bibfnamefont{K.}~\bibnamefont{Sakoda}},
  \emph{\bibinfo{title}{Optical properties of photonic crystals}}
  (\bibinfo{publisher}{Springer-Verlag}, \bibinfo{year}{2001}).

\bibitem[{\citenamefont{Joannopoulos et~al.}(2008)\citenamefont{Joannopoulos,
  Johnson, Winn, and Meade}}]{joannopoulos08a}
\bibinfo{author}{\bibfnamefont{J.~D.} \bibnamefont{Joannopoulos}},
  \bibinfo{author}{\bibfnamefont{S.~G.} \bibnamefont{Johnson}},
  \bibinfo{author}{\bibfnamefont{J.~N.} \bibnamefont{Winn}}, \bibnamefont{and}
  \bibinfo{author}{\bibfnamefont{R.~D.} \bibnamefont{Meade}},
  \emph{\bibinfo{title}{Photonic Crystals, Molding the Flow of Light}}
  (\bibinfo{publisher}{Princeton University Press, Princeton},
  \bibinfo{year}{2008}), \bibinfo{edition}{2nd} ed.

\bibitem[{\citenamefont{Kushwaha et~al.}(1993)\citenamefont{Kushwaha, Halevi,
  Dobrzynski, and Djafari-Rouhani}}]{kushwaha93a}
\bibinfo{author}{\bibfnamefont{M.~S.} \bibnamefont{Kushwaha}},
  \bibinfo{author}{\bibfnamefont{P.}~\bibnamefont{Halevi}},
  \bibinfo{author}{\bibfnamefont{L.}~\bibnamefont{Dobrzynski}},
  \bibnamefont{and}
  \bibinfo{author}{\bibfnamefont{B.}~\bibnamefont{Djafari-Rouhani}},
  \bibinfo{journal}{Phys. Rev. Lett.} \textbf{\bibinfo{volume}{71}},
  \bibinfo{pages}{2022} (\bibinfo{year}{1993}).

\bibitem[{\citenamefont{Gralak et~al.}(2000)\citenamefont{Gralak, Enoch, and
  Tayeb}}]{gralak00a}
\bibinfo{author}{\bibfnamefont{B.}~\bibnamefont{Gralak}},
  \bibinfo{author}{\bibfnamefont{S.}~\bibnamefont{Enoch}}, \bibnamefont{and}
  \bibinfo{author}{\bibfnamefont{G.}~\bibnamefont{Tayeb}}, \bibinfo{journal}{J.
  Opt. Soc. Am. A} \textbf{\bibinfo{volume}{17}}, \bibinfo{pages}{1012}
  (\bibinfo{year}{2000}).

\bibitem[{\citenamefont{Luo et~al.}(2002)\citenamefont{Luo, Johnson,
  Joannopoulos, and Pendry}}]{luo02a}
\bibinfo{author}{\bibfnamefont{C.}~\bibnamefont{Luo}},
  \bibinfo{author}{\bibfnamefont{S.~G.} \bibnamefont{Johnson}},
  \bibinfo{author}{\bibfnamefont{J.~D.} \bibnamefont{Joannopoulos}},
  \bibnamefont{and} \bibinfo{author}{\bibfnamefont{J.~B.}
  \bibnamefont{Pendry}}, \bibinfo{journal}{Physical Review B}
  \textbf{\bibinfo{volume}{65}}, \bibinfo{pages}{201104}
  (\bibinfo{year}{2002}).

\bibitem[{\citenamefont{Smith and Schurig}(2003)}]{smith03a}
\bibinfo{author}{\bibfnamefont{D.~R.} \bibnamefont{Smith}} \bibnamefont{and}
  \bibinfo{author}{\bibfnamefont{D.}~\bibnamefont{Schurig}},
  \bibinfo{journal}{Phys. Rev. Lett.} \textbf{\bibinfo{volume}{90}},
  \bibinfo{pages}{077405} (\bibinfo{year}{2003}).

\bibitem[{\citenamefont{Yang et~al.}(2004)\citenamefont{Yang, Page, Liu, Cowan,
  Chan, and Sheng}}]{page04}
\bibinfo{author}{\bibfnamefont{S.}~\bibnamefont{Yang}},
  \bibinfo{author}{\bibfnamefont{J.}~\bibnamefont{Page}},
  \bibinfo{author}{\bibfnamefont{Z.}~\bibnamefont{Liu}},
  \bibinfo{author}{\bibfnamefont{M.}~\bibnamefont{Cowan}},
  \bibinfo{author}{\bibfnamefont{C.}~\bibnamefont{Chan}}, \bibnamefont{and}
  \bibinfo{author}{\bibfnamefont{P.}~\bibnamefont{Sheng}},
  \bibinfo{journal}{Phys. Rev. Lett.} \textbf{\bibinfo{volume}{93}},
  \bibinfo{pages}{024301} (\bibinfo{year}{2004}).

\bibitem[{\citenamefont{Cervera et~al.}(2002)\citenamefont{Cervera, Sanchis,
  Sanchez-Perez, Martínez-Sala, Rubio, Meseguer, Lopez, Caballero, and
  Sanchez-Dehesa}}]{cervera02}
\bibinfo{author}{\bibfnamefont{F.}~\bibnamefont{Cervera}},
  \bibinfo{author}{\bibfnamefont{L.}~\bibnamefont{Sanchis}},
  \bibinfo{author}{\bibfnamefont{J.}~\bibnamefont{Sanchez-Perez}},
  \bibinfo{author}{\bibfnamefont{R.}~\bibnamefont{Martínez-Sala}},
  \bibinfo{author}{\bibfnamefont{C.}~\bibnamefont{Rubio}},
  \bibinfo{author}{\bibfnamefont{F.}~\bibnamefont{Meseguer}},
  \bibinfo{author}{\bibfnamefont{C.}~\bibnamefont{Lopez}},
  \bibinfo{author}{\bibfnamefont{D.}~\bibnamefont{Caballero}},
  \bibnamefont{and}
  \bibinfo{author}{\bibfnamefont{J.}~\bibnamefont{Sanchez-Dehesa}},
  \bibinfo{journal}{Phys. Rev. Lett.} \textbf{\bibinfo{volume}{88}},
  \bibinfo{pages}{023902} (\bibinfo{year}{2002}).

\bibitem[{\citenamefont{Zhang et~al.}(2009)\citenamefont{Zhang, Yin, and
  Fang}}]{zhang09}
\bibinfo{author}{\bibfnamefont{S.}~\bibnamefont{Zhang}},
  \bibinfo{author}{\bibfnamefont{L.}~\bibnamefont{Yin}}, \bibnamefont{and}
  \bibinfo{author}{\bibfnamefont{N.}~\bibnamefont{Fang}},
  \bibinfo{journal}{Phys. Rev. Lett.} \textbf{\bibinfo{volume}{102}},
  \bibinfo{pages}{194301} (\bibinfo{year}{2009}).

\bibitem[{\citenamefont{Kaina et~al.}(2015)\citenamefont{Kaina, Lemoult, Fink,
  and Lerosey}}]{kaina2015}
\bibinfo{author}{\bibfnamefont{N.}~\bibnamefont{Kaina}},
  \bibinfo{author}{\bibfnamefont{F.}~\bibnamefont{Lemoult}},
  \bibinfo{author}{\bibfnamefont{M.}~\bibnamefont{Fink}}, \bibnamefont{and}
  \bibinfo{author}{\bibfnamefont{G.}~\bibnamefont{Lerosey}},
  \bibinfo{journal}{Nature} \textbf{\bibinfo{volume}{525}}, \bibinfo{pages}{77}
  (\bibinfo{year}{2015}).

\bibitem[{\citenamefont{Veselago}(1968)}]{veselago68a}
\bibinfo{author}{\bibfnamefont{V.~G.} \bibnamefont{Veselago}},
  \bibinfo{journal}{Sov. Phys. Usp.} \textbf{\bibinfo{volume}{10}}
  (\bibinfo{year}{1968}).

\bibitem[{\citenamefont{Pendry}(2000)}]{pendry00a}
\bibinfo{author}{\bibfnamefont{J.~B.} \bibnamefont{Pendry}},
  \bibinfo{journal}{Phys. Rev. Lett.} \textbf{\bibinfo{volume}{85}},
  \bibinfo{pages}{3966} (\bibinfo{year}{2000}).

\bibitem[{\citenamefont{Christensen and de~Abajo}(2012)}]{christensen12}
\bibinfo{author}{\bibfnamefont{J.}~\bibnamefont{Christensen}} \bibnamefont{and}
  \bibinfo{author}{\bibfnamefont{F.}~\bibnamefont{de~Abajo}},
  \bibinfo{journal}{Phys. Rev. Lett.} \textbf{\bibinfo{volume}{108}},
  \bibinfo{pages}{124301} (\bibinfo{year}{2012}).

\bibitem[{\citenamefont{Poddubny et~al.}(2013)\citenamefont{Poddubny, Iorsh,
  Belov, and Kivshar}}]{belov13a}
\bibinfo{author}{\bibfnamefont{A.}~\bibnamefont{Poddubny}},
  \bibinfo{author}{\bibfnamefont{I.}~\bibnamefont{Iorsh}},
  \bibinfo{author}{\bibfnamefont{P.}~\bibnamefont{Belov}}, \bibnamefont{and}
  \bibinfo{author}{\bibfnamefont{Y.}~\bibnamefont{Kivshar}},
  \bibinfo{journal}{Nat Photon} \textbf{\bibinfo{volume}{7}},
  \bibinfo{pages}{948} (\bibinfo{year}{2013}).

\bibitem[{\citenamefont{Bertling et~al.}(2014)\citenamefont{Bertling, Perchoux,
  Taimre, Malkin, Robert, Raki{\'c}, and Bosch}}]{bertling_OE_2014}
\bibinfo{author}{\bibfnamefont{K.}~\bibnamefont{Bertling}},
  \bibinfo{author}{\bibfnamefont{J.}~\bibnamefont{Perchoux}},
  \bibinfo{author}{\bibfnamefont{T.}~\bibnamefont{Taimre}},
  \bibinfo{author}{\bibfnamefont{R.}~\bibnamefont{Malkin}},
  \bibinfo{author}{\bibfnamefont{D.}~\bibnamefont{Robert}},
  \bibinfo{author}{\bibfnamefont{A.~D.} \bibnamefont{Raki{\'c}}},
  \bibnamefont{and} \bibinfo{author}{\bibfnamefont{T.}~\bibnamefont{Bosch}},
  \bibinfo{journal}{Optics express} \textbf{\bibinfo{volume}{22}},
  \bibinfo{pages}{30346} (\bibinfo{year}{2014}).

\bibitem[{\citenamefont{Zipser et~al.}(2003)\citenamefont{Zipser, Franke,
  Olsson, Molin, and Sj{\"o}dahl}}]{zipser_AO_2003}
\bibinfo{author}{\bibfnamefont{L.}~\bibnamefont{Zipser}},
  \bibinfo{author}{\bibfnamefont{H.}~\bibnamefont{Franke}},
  \bibinfo{author}{\bibfnamefont{E.}~\bibnamefont{Olsson}},
  \bibinfo{author}{\bibfnamefont{N.-E.} \bibnamefont{Molin}}, \bibnamefont{and}
  \bibinfo{author}{\bibfnamefont{M.}~\bibnamefont{Sj{\"o}dahl}},
  \bibinfo{journal}{Applied optics} \textbf{\bibinfo{volume}{42}},
  \bibinfo{pages}{5831} (\bibinfo{year}{2003}).

\bibitem[{\citenamefont{Malkin et~al.}(2014)\citenamefont{Malkin, Todd, and
  Robert}}]{malkin_JSV_2014}
\bibinfo{author}{\bibfnamefont{R.}~\bibnamefont{Malkin}},
  \bibinfo{author}{\bibfnamefont{T.}~\bibnamefont{Todd}}, \bibnamefont{and}
  \bibinfo{author}{\bibfnamefont{D.}~\bibnamefont{Robert}},
  \bibinfo{journal}{Journal of Sound and Vibration}
  \textbf{\bibinfo{volume}{333}}, \bibinfo{pages}{4473 }
  (\bibinfo{year}{2014}).

\bibitem[{\citenamefont{Wang and Zunger}(1994)}]{wang1994}
\bibinfo{author}{\bibfnamefont{L.}~\bibnamefont{Wang}} \bibnamefont{and}
  \bibinfo{author}{\bibfnamefont{A.}~\bibnamefont{Zunger}},
  \bibinfo{journal}{J. Chem. Phys.} \textbf{\bibinfo{volume}{100}},
  \bibinfo{pages}{2394} (\bibinfo{year}{1994}).

\bibitem[{\citenamefont{Ho et~al.}(1990)\citenamefont{Ho, Chan, and
  Soukoulis}}]{soukoulis90}
\bibinfo{author}{\bibfnamefont{K.}~\bibnamefont{Ho}},
  \bibinfo{author}{\bibfnamefont{C.}~\bibnamefont{Chan}}, \bibnamefont{and}
  \bibinfo{author}{\bibfnamefont{C.}~\bibnamefont{Soukoulis}},
  \bibinfo{journal}{Phys. Rev. Lett.} \textbf{\bibinfo{volume}{65}},
  \bibinfo{pages}{3152} (\bibinfo{year}{1990}).

\bibitem[{\citenamefont{Johnson and Joannopoulos}(2001)}]{johnson2001}
\bibinfo{author}{\bibfnamefont{S.}~\bibnamefont{Johnson}} \bibnamefont{and}
  \bibinfo{author}{\bibfnamefont{J.}~\bibnamefont{Joannopoulos}},
  \bibinfo{journal}{Opt. Express} \textbf{\bibinfo{volume}{8}},
  \bibinfo{pages}{173} (\bibinfo{year}{2001}).

\bibitem[{\citenamefont{Poulton et~al.}(2000)\citenamefont{Poulton, Movchan,
  McPhedran, Nicorovici, and Antipov}}]{poulton2000}
\bibinfo{author}{\bibfnamefont{C.}~\bibnamefont{Poulton}},
  \bibinfo{author}{\bibfnamefont{A.}~\bibnamefont{Movchan}},
  \bibinfo{author}{\bibfnamefont{R.}~\bibnamefont{McPhedran}},
  \bibinfo{author}{\bibfnamefont{N.}~\bibnamefont{Nicorovici}},
  \bibnamefont{and} \bibinfo{author}{\bibfnamefont{Y.}~\bibnamefont{Antipov}},
  \bibinfo{journal}{Proceedings of the Royal Society of London A}
  \textbf{\bibinfo{volume}{456}}, \bibinfo{pages}{2543} (\bibinfo{year}{2000}).

\bibitem[{\citenamefont{Vanel et~al.}(2016)\citenamefont{Vanel, Craster,
  Colquitt, and Makwana}}]{vanel16a}
\bibinfo{author}{\bibfnamefont{A.~L.} \bibnamefont{Vanel}},
  \bibinfo{author}{\bibfnamefont{R.~V.} \bibnamefont{Craster}},
  \bibinfo{author}{\bibfnamefont{D.~J.} \bibnamefont{Colquitt}},
  \bibnamefont{and} \bibinfo{author}{\bibfnamefont{M.}~\bibnamefont{Makwana}},
  \bibinfo{journal}{Wave Motion} \textbf{\bibinfo{volume}{67}},
  \bibinfo{pages}{15} (\bibinfo{year}{2016}).

\bibitem[{\citenamefont{Kittel}(1996)}]{kittel96a}
\bibinfo{author}{\bibfnamefont{C.}~\bibnamefont{Kittel}},
  \emph{\bibinfo{title}{Introduction to solid state physics}}
  (\bibinfo{publisher}{John Wiley \& Sons}, \bibinfo{address}{New York},
  \bibinfo{year}{1996}), \bibinfo{edition}{7th} ed.

\bibitem[{\citenamefont{Brillouin}(1953)}]{brillouin53a}
\bibinfo{author}{\bibfnamefont{L.}~\bibnamefont{Brillouin}},
  \emph{\bibinfo{title}{Wave propagation in periodic structures}}
  (\bibinfo{publisher}{Dover}, \bibinfo{address}{New York},
  \bibinfo{year}{1953}), \bibinfo{edition}{2nd} ed.

\bibitem[{\citenamefont{Harrison et~al.}(2007)\citenamefont{Harrison, Kuchment,
  Sobolev, and Winn}}]{harrison07a}
\bibinfo{author}{\bibfnamefont{J.~M.} \bibnamefont{Harrison}},
  \bibinfo{author}{\bibfnamefont{P.}~\bibnamefont{Kuchment}},
  \bibinfo{author}{\bibfnamefont{A.}~\bibnamefont{Sobolev}}, \bibnamefont{and}
  \bibinfo{author}{\bibfnamefont{B.}~\bibnamefont{Winn}}, \bibinfo{journal}{J.
  Phys. A - Math} \textbf{\bibinfo{volume}{40}}, \bibinfo{pages}{7597}
  (\bibinfo{year}{2007}).

\bibitem[{\citenamefont{Adams et~al.}(2008)\citenamefont{Adams, Craster, and
  Guenneau}}]{adams08a}
\bibinfo{author}{\bibfnamefont{S.~D.~M.} \bibnamefont{Adams}},
  \bibinfo{author}{\bibfnamefont{R.~V.} \bibnamefont{Craster}},
  \bibnamefont{and} \bibinfo{author}{\bibfnamefont{S.}~\bibnamefont{Guenneau}},
  \bibinfo{journal}{Proc. R. Soc. Lond. A} \textbf{\bibinfo{volume}{464}},
  \bibinfo{pages}{2669} (\bibinfo{year}{2008}).

\bibitem[{\citenamefont{Craster et~al.}(2012)\citenamefont{Craster,
  Antonakakis, Makwana, and Guenneau}}]{craster12a}
\bibinfo{author}{\bibfnamefont{R.~V.} \bibnamefont{Craster}},
  \bibinfo{author}{\bibfnamefont{T.}~\bibnamefont{Antonakakis}},
  \bibinfo{author}{\bibfnamefont{M.}~\bibnamefont{Makwana}}, \bibnamefont{and}
  \bibinfo{author}{\bibfnamefont{S.}~\bibnamefont{Guenneau}},
  \bibinfo{journal}{Phys. Rev. B} \textbf{\bibinfo{volume}{86}},
  \bibinfo{pages}{115130} (\bibinfo{year}{2012}).

\bibitem[{\citenamefont{Craster et~al.}(2010)\citenamefont{Craster, Kaplunov,
  and Pichugin}}]{craster10a}
\bibinfo{author}{\bibfnamefont{R.~V.} \bibnamefont{Craster}},
  \bibinfo{author}{\bibfnamefont{J.}~\bibnamefont{Kaplunov}}, \bibnamefont{and}
  \bibinfo{author}{\bibfnamefont{A.~V.} \bibnamefont{Pichugin}},
  \bibinfo{journal}{Proc R Soc Lond A} \textbf{\bibinfo{volume}{466}},
  \bibinfo{pages}{2341} (\bibinfo{year}{2010}).

\bibitem[{\citenamefont{Bosch et~al.}(2005)\citenamefont{Bosch, Bes, Scalize,
  and Plantier}}]{Bosch_EOS_05}
\bibinfo{author}{\bibfnamefont{T.}~\bibnamefont{Bosch}},
  \bibinfo{author}{\bibfnamefont{C.}~\bibnamefont{Bes}},
  \bibinfo{author}{\bibfnamefont{L.}~\bibnamefont{Scalize}}, \bibnamefont{and}
  \bibinfo{author}{\bibfnamefont{G.}~\bibnamefont{Plantier}},
  \bibinfo{journal}{Encyclopedia of Sensors} \textbf{\bibinfo{volume}{X}},
  \bibinfo{pages}{1} (\bibinfo{year}{2005}).

\bibitem[{\citenamefont{Taimre et~al.}(2015)\citenamefont{Taimre, Nikoli{\'c},
  Bertling, Lim, Bosch, and Raki{\'c}}}]{taimre_AOP_15}
\bibinfo{author}{\bibfnamefont{T.}~\bibnamefont{Taimre}},
  \bibinfo{author}{\bibfnamefont{M.}~\bibnamefont{Nikoli{\'c}}},
  \bibinfo{author}{\bibfnamefont{K.}~\bibnamefont{Bertling}},
  \bibinfo{author}{\bibfnamefont{Y.~L.} \bibnamefont{Lim}},
  \bibinfo{author}{\bibfnamefont{T.}~\bibnamefont{Bosch}}, \bibnamefont{and}
  \bibinfo{author}{\bibfnamefont{A.~D.} \bibnamefont{Raki{\'c}}},
  \bibinfo{journal}{Advances in Optics and Photonics}
  \textbf{\bibinfo{volume}{7}}, \bibinfo{pages}{570} (\bibinfo{year}{2015}).

\bibitem[{\citenamefont{Roumy et~al.}(2015)\citenamefont{Roumy, Perchoux, Lim,
  Taimre, Raki\'{c}, and Bosch}}]{AlRoumy_AO_15}
\bibinfo{author}{\bibfnamefont{J.~A.} \bibnamefont{Roumy}},
  \bibinfo{author}{\bibfnamefont{J.}~\bibnamefont{Perchoux}},
  \bibinfo{author}{\bibfnamefont{Y.~L.} \bibnamefont{Lim}},
  \bibinfo{author}{\bibfnamefont{T.}~\bibnamefont{Taimre}},
  \bibinfo{author}{\bibfnamefont{A.~D.} \bibnamefont{Raki\'{c}}},
  \bibnamefont{and} \bibinfo{author}{\bibfnamefont{T.}~\bibnamefont{Bosch}},
  \bibinfo{journal}{Appl. Opt.} \textbf{\bibinfo{volume}{54}},
  \bibinfo{pages}{312} (\bibinfo{year}{2015}).

\bibitem[{\citenamefont{Scruby and Drain}(1990)}]{scruby1990}
\bibinfo{author}{\bibfnamefont{C.~B.} \bibnamefont{Scruby}} \bibnamefont{and}
  \bibinfo{author}{\bibfnamefont{L.~E.} \bibnamefont{Drain}},
  \emph{\bibinfo{title}{Laser ultrasonics techniques and applications}}
  (\bibinfo{publisher}{CRC Press}, \bibinfo{year}{1990}).

\end{thebibliography}

\end{document}